\documentclass[aps, prd, twocolumn, superscriptaddress, nofootinbib]{revtex4-1}

\usepackage{graphicx}% Include figure files
\usepackage{dcolumn}% Align table columns on decimal point
\usepackage{bm}% bold math
\usepackage{amsmath}
\usepackage{amssymb}
\usepackage[normalem]{ulem}

\newcommand{\tr}{\ensuremath{\mathrm{Tr}}}
\newcommand{\mbf}[1]{\ensuremath{\boldsymbol{#1}}}

\begin{document}

\title{Color flux tubes in $SU(3)$ Yang-Mills theory: \\ an investigation with
the connected correlator}

\author{Nico Battelli}
\email{battelln@tcd.ie}
\altaffiliation{Present adress:
School of Mathematics, Trinity College Dublin, Dublin 2, Ireland}
\affiliation{Universit\`a di Pisa and INFN Sezione di Pisa,\\ 
Largo Pontecorvo 3, I-56127 Pisa, Italy
}

\author{Claudio Bonati}
\email{claudio.bonati@df.unipi.it}
\affiliation{Universit\`a di Pisa and INFN Sezione di Pisa,\\ 
Largo Pontecorvo 3, I-56127 Pisa, Italy
}

\date{\today}% It is always \today, today,
             % but any date may be explicitly specified

\begin{abstract}
In this work we perform an investigation of the flux tube between two static
color sources in four dimensional $SU(3)$ Yang-Mills theory, using the so called connected
correlator.  Contrary to most previous studies we do not use any smoothing
algorithm to facilitate the evaluation of the correlator, that is performed
using only stochastically exact techniques. We first examine the
renormalization properties of the connected operator, then we present our
numerical data for the longitudinal chromoelectric component of the flux tube,
that are used to extract the dual superconductivity parameters. 
\end{abstract}

\maketitle

\section{Introduction}\label{sec:intro}

The investigation of the color flux tubes connecting static sources in
non-abelian gauge theories has become a standard tool to study color
confinement \cite{Fukugita:1983du, Flower:1985gs, Wosiek:1987kx, Sommer:1987uz,
DiGiacomo:1989yp, DiGiacomo:1990hc, Bali:1994de, Haymaker:1994fm, Cea:1995zt,
Okiharu:2003vt, Bissey:2006bz, Bicudo:2011hk, Bakry:2014gea}.  Indeed, in
lattice simulations, color sources are seen to be connected by tube-like
structures for all the values of the coupling constant (at zero temperature).
This is a strong indication that the mechanism responsible for color
confinement is the same at weak and at strong coupling, in which limit color flux tubes
naturally emerge \cite{Kogut:1974ag}, and the area-law of the Wilson loops can
be analytically proven \cite{Osterwalder:1977pc}.

To study flux tubes on the lattice we need an observable whose average value
will provide us information about the flux tube details: more specifically the
value of this observable has to be related to that of the field strength in the
background of a couple of static color sources. A quantity that satisfies
this requirement can be built by using the correlator of a Polyakov loops pair
with a plaquette: the pair of Polyakov loops represents a couple of static
color charges (a Wilson loop is also often used for this purpose) while the
plaquette probes the field strength in the background of the static charges. 

\begin{figure}[t] 
\centering 
\includegraphics[width=0.8\columnwidth, clip]{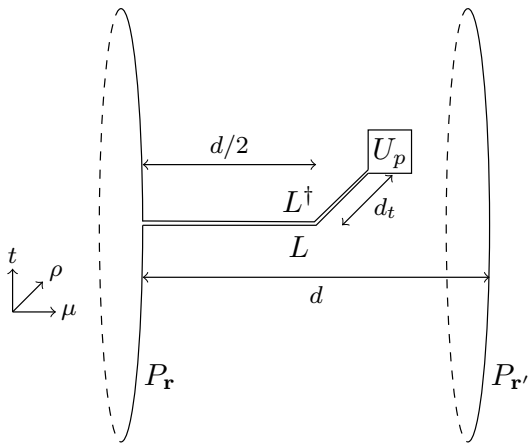}
\caption{Graphical representation of the numerator of the first term of
Eq.~\eqref{eq:conn_corr}. Ellipses denote the Polyakov loops (at distance
$d$ from each other) and the string $L$ connects one Polyakov loop with the
plaquette, first reaching the midpoint between the Polyakov loops at $d/2$,
then moving in the transverse direction for a distance $d_t$. The plaquette is
drawn parallel to the plane identified by the two Polyakov loops since this is
the case that will be studied in this paper, but its orientation can \emph{a
priori} be generic.} 
\label{fig:rho_conn}
\end{figure}

This general idea is common to all numerical implementations, however in the
literature two different ways of defining the basic correlator are present: the
first possibility is to use the expression \cite{Fukugita:1983du}  
\begin{equation}\label{eq:disc_corr}
\rho_{disc}=\frac{\langle \tr(P_{\mbf{r}})\tr(P_{\mbf{r}'}^{\dag})\tr(U_p)\rangle}{
\langle \tr(P_{\mbf{r}})\tr(P_{\mbf{r}'}^{\dag}))\rangle}-\langle \tr(U_p)\rangle\ ,
\end{equation} 
where $P_{\mbf{r}}$ stands for the Polyakov loop at spatial position $\mbf{r}$
and $U_p$ for the plaquette operator; this is known as the ``disconnected''
correlator. Another possibility is to use the definition
\cite{DiGiacomo:1989yp, DiGiacomo:1990hc}
\begin{equation}\label{eq:conn_corr}
\begin{gathered}
\rho_{conn}=\frac{\langle \tr(P_{\mbf{r}}LU_pL^{\dag})\tr(P_{\mbf{r}'}^{\dag})\rangle}{\langle\tr(P_{\mbf{r}})\tr(P_{\mbf{r}'}^{\dag})\rangle} - \\ 
-\frac{1}{N_c}\frac{\langle\tr(P_{\mbf{r}})\tr(P_{\mbf{r}'}^{\dag})\tr(U_p)\rangle}{\langle\tr(P_{\mbf{r}})\tr(P_{\mbf{r}'}^{\dag})\rangle}\ ,
\end{gathered}
\end{equation}
where $N_c$ is the number of colors and $L$ is the parallel transporter
associated to the path shown in Fig.~\ref{fig:rho_conn}; $\rho_{conn}$ is known
as the ``connected'' correlator and $L$ is often called the ``Schwinger line''
in the literature.  Both $\rho_{disc}$ and $\rho_{conn}$ are related to color
flux tubes but they are not equivalent; in fact they are associated to different physical
observables. This can be readily understood by looking at their \emph{naive}
continuum limit, in which the plaquette operator $U_p$ is expanded in powers of
the lattice spacing $a$: $\rho_{disc}$ scales to the continuum as $a^4$ and
gives access to\footnote{We assume the gauge group to be $SU(N_c)$. The case of
abelian groups is somehow exceptional in the present context, since
$\rho_{disc}$ is linear in the field strength for abelian groups.}
$\tr(F_{\mu\nu}^2)$ (we are assuming $U_p$ oriented in the $\mu\nu$ plane),
$\rho_{conn}$ scales to the continuum as $a^2$ and it is linear in $F_{\mu\nu}$
(see e.g. \cite{DiGiacomo:1989yp, DiGiacomo:1990hc} for more details).

Since $\rho_{disc}$ and $\rho_{conn}$ do not provide equivalent physical
information, the choice of the operator to be used requires some discussion.
Two arguments that have been adopted in the past to advocate the use of one
operator or the other are the following: on one hand the operator $\rho_{disc}$
is theoretically better understood, since it can be easily shown to be
multiplicatively renormalizable (see the discussion in Sec.~\ref{sec:ren}), and
the renormalization constant needed to cancel logarithmic divergences can be
fixed by using a lattice sum rule (see e.g.  \cite{Bali:1994de}). On the other
hand $\rho_{disc}$ is noisier than $\rho_{conn}$, and noise reduction was the original
motivation for the introduction of the connected correlator in
\cite{DiGiacomo:1989yp, DiGiacomo:1990hc}: since $\rho_{disc}$ probes the
square of the field-strength it is more sensitive to ultraviolet (UV) fluctuations. 

The choice of the operator to be used was thus largely based on the importance
attributed to fluctuations. Sometimes UV fluctuations have a prominent role in
the physical phenomenon to be studied, a prototypical example being the
fluctuation-induced broadening of flux tubes \cite{Luscher:1980iy}. In these
cases the choice of the disconnected operator is mandatory, and specific
stochastically exact noise-reduction techniques have been typically adopted to measure it
\cite{Allais:2008bk, Gliozzi:2010zv, Gliozzi:2010jh, Cardoso:2013lla,
Amado:2013rja, Caselle:2016mqu}. 

When fluctuations were not expected to be important for the physical problem
studied, the operator $\rho_{conn}$ has been the most common choice
\cite{Cardaci:2010tb, Cea:2012qw, Cea:2014uja, Cea:2015wjd, Baker:2018mhw},
supplemented by the use of smoothing algorithms to reduce UV noise. In studies
performed with dynamical fermions only $\rho_{conn}$ has been used so far
\cite{Cea:2017ocq, Bonati:2018uwh}, since the accessible statistics are much
lower than in the pure glue case, and stochastically exact error reduction
techniques (see \cite{Ce:2016ajy}) are not easily applicable and still not widely used.

An important point to be noted is that smoothing has always been used to reduce
the effect of UV fluctuations in $\rho_{conn}$, however 
this standard procedure can \emph{a priori} also induce systematical errors in
the flux tube measure. Indeed in \cite{Bonati:2018uwh} a worrisome dependence
of the physical results on the amount of smoothing adopted was noted
(see also \cite{Cea:2012qw} for the case of Yang-Mills theory).

The aim of this work is to study $\rho_{conn}$ without using any smoothing
algorithm, in order to understand if the connected operator can be used in a
coherent field-theoretical setup to extract physical quantities related to flux
tubes.  For this purpose we evaluate the connected correlator $\rho_{conn}$ in
four dimensional $SU(3)$ Yang-Mills theory using only stochastically exact
techniques (i.e.  multihit \cite{Parisi:1983hm} and multilevel
\cite{Luscher:2001up} algorithms).  To physically interpret these data we need
to study the renormalization of the connected correlator $\rho_{conn}$: in our
data the singularities related to the continuum limit (that are usually hidden
by the use of smoothing) are clearly visible and we need to take care of them.
The renormalization of $\rho_{conn}$ is far less trivial than that of
$\rho_{disc}$, however we will show that $\rho_{conn}$ renormalizes
multiplicatively, and it can be used to extract physically relevant
information.

The paper is organized as follows: in Sec.~\ref{sec:ren} we discuss the issues
related to the renormalization of $\rho_{conn}$, using arguments largely based
on \cite{Berwein:2012mw, Berwein:2013xza}, where the renormalization of cyclic
Wilson loops was addressed.  In Sec.~\ref{sec:numres} we introduce the
numerical setup adopted, we present the results obtained for the longitudinal
chromoelectric field, and we discuss the physical implications of these results
for the dual superconductor model of the vacuum. Finally, in
Sec.~\ref{sec:concl} we draw our conclusions.

\section{Renormalization of $\rho_{conn}$}\label{sec:ren}

In order to discuss the renormalization of $\rho_{conn}$ it seems appropriate
to start by briefly recalling some general facts about the renomalization of loop
operators.

A loop operator is a generalized Wilson loop, which in
the continuum can be written as
\begin{equation}\label{eq:loop_op}
W_C=\tr\left[ \mathcal{P} \exp\left(i\oint_C A_{\mu}\mathrm{d} x^{\mu}\right)\right] ,
\end{equation} 
where $C$ is a closed curve and $\mathcal{P}$ stands for path-ordered. The systematic
study of the divergences associated to these operators in four dimensional
gauge theory was initiated in \cite{Polyakov:1980ca}, where it was suggested
that $W_C$ is multiplicatively renormalizable if $C$ is piecewise smooth and
not self-intersecting. From a one loop computation with cut-off regularization
two sources of divergences were identified in \cite{Polyakov:1980ca}:
logarithmic divergences originate from the points at which $C$ is not
differentiable, while linear divergences are always present, they exponentiate
and globally contribute with a term of the form $\exp(c\mathcal{L}(C)/a)$,
where $\mathcal{L}(C)$ is the length of the curve, $a$ is the UV cut-off and $c$
is a constant.

If $C$ is smooth it was shown in \cite{Dotsenko:1979wb} that $W_C$ is finite at
all orders of perturbation theory, after the usual charge renormalization is
performed. The case of a non-smooth curve was studied in \cite{Brandt:1981kf},
where it was proven that to each cusp with angle $\gamma$ a multiplicative
renormalization constant has to be associated, whose value depends just on
$\gamma$. The case of self-intersecting curves was also studied in
\cite{Brandt:1981kf}: in this case operator mixing between operators
corresponding to different color contractions at the crossing points has also
to be taken into account. The final result is the following: if $r$
intersection points (corresponding to the sets of intersection angles
$\{\theta_1\},\ldots,\{\theta_r\}$) and $s$ cusps (corresponding to the angles
$\gamma_1,\ldots,\gamma_s$) are present, then renormalization matrices and
renormalization constants exist such that every color contraction
$W_C^{i_1,\ldots,i_r}$ can be renormalized by using 
\begin{equation}\label{eq:generalZnoexp}
\begin{aligned}
\left.W_C^{i_1,\ldots,i_r}\right|_R &=Z(\gamma_1)\cdots Z(\gamma_s)\times \\ 
& \times Z_{i_1j_1}(\{\theta_1\})\cdots Z_{i_rj_r}(\{\theta_r\}) W_C^{j_1,\ldots,j_r}\ ,
\end{aligned}
\end{equation}
where the exponentials associated to linear divergences are implied.  In
\cite{Brandt:1981kf} the possibility for different color contractions to have
different lengths was however not considered, and this could seem to be a
source of problems for the renormalization of $\rho_{conn}$.

Let us start by studying the renormalization of $\rho_{conn}$ in a scheme in
which no power-law divergences are present (like e.g.  minimal subtraction), so
that we can use Eq.~\eqref{eq:generalZnoexp} without worrying of the
complications related to linear divergences. In this case Polyakov loops do not
need any renormalization, since they are associated to smooth contours, and the
denominator $\langle\tr(P_{\mbf{r}})\tr(P_{\mbf{r}'}^{\dag})\rangle$ of
Eq.~\eqref{eq:conn_corr} is finite once charge renormalization is performed.
For the same reason the term $\tr(P_{\mbf{r}'}^{\dag})$ in the numerator is
also harmless and we can just concentrate on the term 
\begin{equation}\label{eq:O1}
O_1=\tr\left(P_{\mbf{r}}LU_pL^{\dag}\right)\ .
\end{equation}
The path associated to $O_1$ is not smooth at three point: the point at which
$L$ and $L^{\dag}$ connects to $P_{\mbf{r}}$, the point at which they connects
to $U_p$ and the corner point of $L$ and $L^{\dag}$. All the rest of the
contour contributes only to linear divergences, that we are neglecting for the moment.

In studying the renormalization of $O_1$ we have to take into account the
mixing with all the operators that can be build from $O_1$ using different
color contractions at the crossing points. Eight color contractions can be
built (2 different contractions for each crossing point), however it is easy to
realize, using $LL^{\dag}=1$ and analogous relations, that all the contractions
that are not equal to $O_1$ are equal to
\begin{equation}\label{eq:O2}
O_2=\frac{1}{N_c}\tr\left(P_{\mbf{r}}\right)\tr\left(U_p\right)\ ,
\end{equation}
where the $1/N_c$ factor is needed to keep the same normalization of
Eq.~\eqref{eq:O1}. From the previous general discussion it follows immediately that
$O_2$ is multiplicatively renormalizable, a fact that will be used soon.

Eq.~\eqref{eq:generalZnoexp} implies that we have to consider in general an
$8\times 8$ mixing matrix, which is the tensor product of the three basic
$2\times 2$ mixing matrices, however very stringent constraints are imposed
on this $8\times 8$ matrix by the fact that only two color contractions
are different from each other, and by the fact that $O_2$ is multiplicatively
renormalizable. 

Let us discuss explicitly the case $d_t=0$ (see Fig.~\ref{fig:rho_conn}), in
which we have just the tensor product of two $2\times 2$ mixing matrices and
$4$ color contractions. Denoting the two $2\times 2$ mixing matrices by $Z^A$
and $Z^B$ we thus have\footnote{For the sake of the simplicity we do not show
explicitly the cusp renormalization factors $Z(\gamma_1), \ldots, Z(\gamma_s)$,
that multiplies all the lines of the right hand side of the following
equation.} 
\begin{equation}\label{eq:mulrenproof}
\left(\begin{array}{c} O_1^{(R)} \\ O_2^{(R)} \\ O_2^{(R)} \\ O_2^{(R)}\end{array}\right)
=\left(\begin{array}{cc} Z^A_{00} Z^B & Z^A_{01}Z^B \\
Z^A_{10}Z^B & Z^A_{11}Z^B \end{array}\right)
\left(\begin{array}{c} O_1 \\ O_2 \\ O_2 \\ O_2\end{array}\right)\ ,
\end{equation}
where the apex $(R)$ stands for ``renormalized'' and 
\begin{equation}
Z^B=\left(\begin{array}{cc} Z^B_{00} & Z^B_{01} \\ 
Z^B_{10} & Z^B_{11}\end{array}\right)\ .
\end{equation}
Since $O_2$ renormalizes multiplicatively, the coefficients $Z^A_{10}$ and
$Z^B_{10}$ have to vanish, otherwise $O_2^{(R)}$ would depend also on $O_1$. If
we use $Z^A_{10}=Z^B_{10}=0$ we see that the last three lines of
Eq.~\eqref{eq:mulrenproof} are consistent with each other only if
\begin{equation}
Z^A_{00}+Z^A_{01}=Z^A_{11}\ ,\quad
Z^B_{00}+Z^B_{01}=Z^A_{11}\ .
\end{equation}
It is then simple to verify that Eq.~\eqref{eq:mulrenproof} collapses to
\begin{equation}
\left(\begin{array}{c} O_1^{(R)} \\ O_2^{(R)}\end{array}\right)=
\left(\begin{array}{cc} Z_1 & Z_2-Z_1 \\ 0 & Z_2\end{array}\right)
\left(\begin{array}{c} O_1 \\ O_2\end{array}\right)\ ,
\end{equation} 
where 
\begin{equation}\label{eq:Z_1}
Z_1=Z^A_{00}Z^{B}_{00}\ ,\quad Z_2=Z^A_{11}Z^B_{11}\ .
\end{equation}
As a consequence we finally have
\begin{equation}
O_1^{(R)}-O_2^{(R)}=Z_1(O_1-O_2)
\end{equation}
which means that $\rho_{conn}$ renormalizes
multiplicatively. 

The same argument (which is an adaptation of the one used in
\cite{Berwein:2012mw}) can be repeated without changes also when $d_t>0$, in
which case we have to start from the tensor product of three $2\times 2$ mixing
matrices.  Since the renormalization constants only depend on the set of
intersection angles, the value of $Z_1$ is the same for all the positive $d_t$
values, but it differs from the one at $d_t=0$, due to the presence of new
logarithmic divergences for $d_t>0$.

Let us now consider a renormalization scheme in which linear divergences are
present. The argument to show that $\rho_{conn}$ is multiplicatively
renormalizable also in this case is exactly the same that was used for cyclic
Wilson loops in \cite{Berwein:2013xza}, to which we refer for further details
and some enlightening examples. Here we will just briefly sketch some basic
steps of the proof for the benefit of the reader and to fix the notation. The
main ingredient that is needed is the relation between an operator acting in
the fundamental representation of $SU(N_c)$ (that we will denote by $U$) and
the corresponding operator acting in the adjoint representation (that we will
denote by $U^{\mathrm{adj}}$), which is
\begin{equation}\label{eq:fundadj}
U^{\mathrm{adj}}_{ab}=2\tr(U^{\dag}T_aUT_b)\ ,
\end{equation}
where $T_a$ are the $SU(N_c)$ generators, with the normalization
$\tr(T_aT_b)=\frac{1}{2}\delta_{ab}$. 

By writing $P_{\mbf{r}}$ and $U_p$ in the base $\{1,T_a\}$ and using
Eq.~\eqref{eq:fundadj} it is simple to show that
\begin{equation}
\begin{aligned}
O_1-O_2&=\tr(P_{\mbf{r}}LU_pL^{\dag})-\frac{1}{N_c}\tr(U_p)\tr(P_{\mbf{r}})=\\
&=2\sum_{ab}\tr(P_{\mbf{r}}T_a)L_{ab}^{\mathrm{adj}}\tr(U_p T_b)\ .
\end{aligned}
\end{equation}
We thus see that $O_1-O_2$ can be written as a single generalized loop function
in which traces in different representations are present: two traces in the
fundamental representation (explicitly denoted by $\tr$) and a trace in the
adjoint representation (the summation on $a,b$). 

This expression is particularly convenient since it was shown in
\cite{Berwein:2013xza} (using the exponentiation theorems of
 \cite{Gardi:2010rn, Gardi:2013ita}) that linear divergences factorize
also in untraced loop operators, and they can be cancelled by multiplicative
factors of the form $\exp(-c_r\mathcal{L}(C_r)/a)$, where $c_r$ is a
representation dependent coefficient, $a$ is the UV cut-off, and
$\mathcal{L}(C_r)$ is the length in physical units of the curve $C_r$
associated to the representation $r$. 

Applying this result to the operator $O_1-O_2$ we get
\begin{equation}
\begin{aligned}
&[O_1-O_2]^{(R)}=\exp\left[-\frac{c_a}{a}\left(\frac{d}{2}+d_t\right)\right]\times\\
&\quad \times \exp\left[-\frac{c_f}{a}L_t-4c_f\right] Z_1 [O_1-O_2]\ ,
\end{aligned}
\end{equation}
where $c_f$ and $c_a$ are the constants associated to the fundamental and
adjoint representations, $\frac{d}{2}+d_t$ is the length of the contour in the
adjoint representation (see Fig.~\ref{fig:rho_conn}) and $L_t$ is the temporal
extent of the lattice in physical units. $Z_1$ is the renormalization constant
associated to logarithmic divergences that was previously introduced in
Eq.~\eqref{eq:Z_1}, while the term $4c_f$ is associated to the plaquette, that
has length $4a$, does not generate linear divergences in the continuum and can
be safely neglected. Since a factor $e^{-c_fL_t/a}$ is associated to each
Polyakov loop, we finally have
\begin{equation}\label{eq:ren}
\rho_{conn}^{(R)}= \exp\left[-\frac{c_a}{a}\left(\frac{d}{2}+d_t\right)\right]Z_1(a, d_t) \rho_{conn}\ ,
\end{equation}
where $Z_1$ is independent of $d$ and assumes two different values for $d_t=0$ or 
$d_t\neq 0$.

\section{Numerical results}\label{sec:numres}

\subsection{Setup}\label{sec:setup}

\begin{table}[t]
\begin{tabular}{|l|l|l|l|l|}
\hline $N_s$ & $\beta$   & $a(\beta)$ [fm]   & $d$ & $n_{\mathrm{upd}}$\\ \hline
20         &  6.0      & 0.0931            &   4 & 200   \\ \hline  
28         &  6.2601   & 0.0621            &   6 & 1000  \\ \hline
36         &  6.47466  & 0.0456            &   8 & 3000  \\ \hline
\end{tabular}
\caption{Simulation parameters: $N_s$ is the lattice extent in lattice units,
$\beta$ is the bare coupling constant, $a(\beta)$ is the lattice spacing, $d$
the distance between the Polyakov loops in lattice units, and $n_{\mathrm{upd}}$
the number of internal updates of the multilevel
algorithm.}\label{tab:simpointrho}
\end{table}

As anticipated in the introduction, we study $\rho_{conn}$ in four dimensional
$SU(3)$ Yang-Mills theory, discretized on the lattice by using the standard
Wilson action \cite{Wilson:1974sk}:
\begin{equation}
S=\sum_p \beta\left(1-\frac{1}{3}\mathrm{Re}\tr U_p\right)\ .
\end{equation}

The connected correlator defined in Eq.~\eqref{eq:conn_corr} can be estimated
by using a combination of multihit \cite{Parisi:1983hm} and multilevel
\cite{Luscher:2001up} techniques: multihit can be applied to the links of the
Polyakov loops and to a single link of the plaquette. The multilevel method can
be easily adapted to measure $\rho_{conn}$, since almost all the links entering
$LU_pL^{\dag}$ are associated to a single time slice of the lattice, with the
only possible exception of three links entering $U_p$ (if the plaquette is a
temporal one). The time slice identified by $L$ has been chosen to lay in the
bulk of the corresponding slice of the multilevel algorithm, so that the links
entering $L$ and $L^{\dag}$ are updated in the multilevel. If a slice of
thickness $\Delta=2a$ was used for the case of a temporal plaquette, the
upper link of the plaquette would thus be fixed during the multilevel. However
for all the cases studied in this work a single level of the multilevel
algorithm was used, with slices of thickness $\Delta=4a$, so that all the
links of $LU_pL^{\dag}$ are updated. The number of internal updates of the
multilevel algorithm was optimized by using a technique analogous to the one
discussed in \cite{Luscher:2001up}.

Simulations were performed on symmetric lattices ($N_s=N_t$), and in
Tab.~\ref{tab:simpointrho} we report the simulation details. The plaquette
$U_p$ which appears in Eq.~\eqref{eq:conn_corr} was always chosen to be
parallel to the plane identified by the two Polyakov loops, since we are
interested in studying the longitudinal chromoelectric field, which in all
previous studies was shown to be the dominant component of the flux tube.

The coupling values were chosen in such a way that $d=4a, 6a$ and $8a$
correspond to the same distance in physical units; for this purpose the
parametrization of $a(\beta)$ obtained in \cite{Necco:2001xg} was used.
Parameters in Tab.~\ref{tab:simpointrho} fix the physical distance between the
Polyakov loops to about $0.37\,\mathrm{fm}$ (using $r_0\simeq 0.5\,\mathrm{fm}$
for the Sommer scale \cite{Sommer:1993ce}), and the dimensionless lattice size
was rescaled in order to have (almost) constant physical volume. Note
that $d\simeq 0.37\,\mathrm{fm}$ is quite smaller than the typical values that
have been used in recent works, which range from $0.54\,\mathrm{fm}$ to
0.$76\,\mathrm{fm}$ (see e.g. \cite{Cea:2014uja, Cea:2015wjd}). Test
simulations were also performed on the lattice $28^4$ with coupling
$\beta=6.0$, which excluded the presence of sizable finite size effects in our
data. Data points corresponding to different values of $d_t$ have been
extracted from independent simulations, hence they are statistically
independent from each other.

\subsection{Results for the chromoelectric field}\label{sec:chfield}

\begin{figure}[t] 
\centering 
\includegraphics[width=0.95\columnwidth, clip]{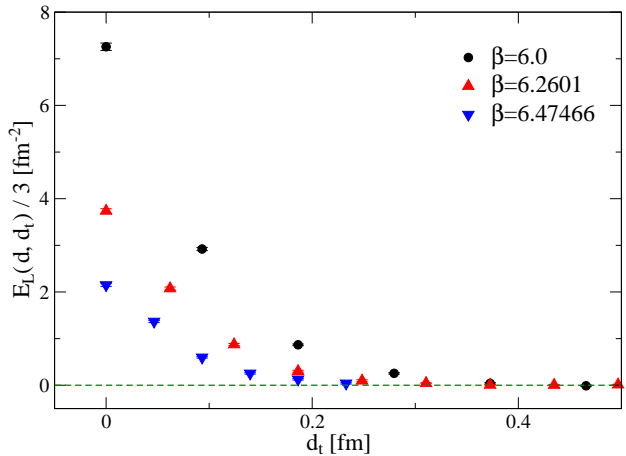}
\caption{$E_L(d, d_t)$ computed from Eq.~\eqref{eq:el_naive} for three values
of the coupling constant and $d\simeq 0.37\,\mathrm{fm}$. }
\label{fig:el_noz}
\end{figure}

From the \emph{naive} continuum limit of $\rho_{conn}$ in
Eq.~\eqref{eq:conn_corr} we can define the longitudinal (due to our choice of
the plaquette orientation, see Sec.~\ref{sec:setup}) chromoelectric field by
using the expression
\begin{equation}\label{eq:el_naive}
E_L(d, d_t)=\frac{\sqrt{\beta/6}}{a^2(\beta)}\,\rho_{conn}(d,d_t)\ ,
\end{equation}
where in our simulations $d$ is fixed to about $0.37\,\mathrm{fm}$, and $d_t$
denotes the transverse distance from the center of the flux tube (see
Fig.~\ref{fig:rho_conn}).  From the discussion in Sec.~\ref{sec:ren} it follows
that we can not expect $E_L$ defined in this way to have a nontrivial
continuum limit. Indeed data shown in Fig.~\ref{fig:el_noz} indicate that
$E_L(d,d_t)$ converges to zero as the continuum limit is approached.

To properly define the continuum limit of $E_L$ we have to use
Eq.~\eqref{eq:ren} and define
\begin{equation}\label{eq:elr}
E_L^{(R)}(d,d_t)= Z_L(d, d_t, a)Z_1(a, d_t) E_L(d, d_t)\ ,
\end{equation}
where $Z_1(a, d_t)$ is the renormalization constant associated to logarithmic
divergences (which is different for $d_t=0$ and $d_t\neq 0$), and we introduced
the shorthand 
\begin{equation}
Z_L(d, d_t, a)=\exp\left[-\frac{c_a}{a}\left(\frac{d}{2}+d_t\right)\right]
\end{equation}
to denote the multiplicative factor needed to remove linear divergences. To
completely define $E_L^R$ we thus have to fix the three constants $c_a$ and
$Z_1(a, d_t)$ (for $d_t=0$ and $d_t\neq 0$). 

The numerical value of $Z_1(a, d_t)$ could be computed in perturbation theory,
however some interesting physical observables can be studied also without a
precise knowledge of this renormalization constant. This is due to the fact that,
for $d_t>0$, $Z_1$ is a multiplicative factor independent of $d_t$, hence the
functional form of $E_L^{(R)}(d, d_t)$ for $d_t>0$ is completely fixed
also without any knowledge of $Z_1$. However, also to study just the functional
form of $E_L^{(R)}(d, d_t)$, we need to fix $c_a$. 

\begin{table}[t]
\begin{tabular}{|l|l|l|}
\hline $N_t$ & $N_s$  & $\beta$ \\ \hline 
4  & 16 &   5.8     \\ \hline   
5  & 20 &   5.91225 \\ \hline
6  & 24 &   6.01388 \\ \hline
7  & 28 &   6.10767 \\ \hline
8  & 32 &   6.19513 \\ \hline
9  & 36 &   6.27708 \\ \hline
10 & 40 &   6.35394 \\ \hline
\end{tabular}
\caption{Simulation points used to fix the value of
the renormalization constant $c_a$}\label{tab:simpointpoly}
\end{table}

Since $c_a$ is a fundamental property of the discretization adopted,
independent of the specific adjoint loop function used and of the infrared
properties of the theory, we have the freedom of choosing the simplest
numerical setup available to fix its value. We decided to extract if from
the continuum scaling of the Polyakov loop in the adjoint representation at
finite temperature, which is an observable that is easily computed to high
precision. 

\begin{figure}[t] 
\centering 
\includegraphics[width=0.9\columnwidth, clip]{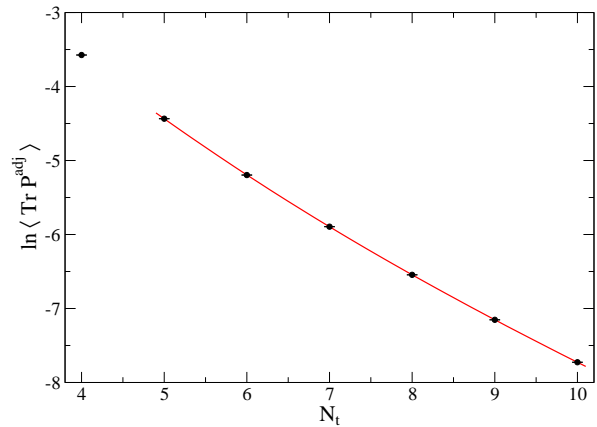}
\caption{Continuum scaling of $\langle \tr P^{\mathrm{adj}}\rangle$ in
the high temperature phase. The continuous line is the result of a fit of the form
$k_0+c_a N_t + k_1/N_t$.}
\label{fig:Z}
\end{figure}

For this purpose we performed simulations starting from a $4\times 16^3$
lattice at $\beta=5.8$ (the deconfinement transition on $N_t=4$ lattices takes
place at $\beta_c=5.6925(2)$, see \cite{Fingberg:1992ju}), then increasing the
value of $N_t$ keeping the physical temperature constant and the aspect ratio
fixed to $4$, see Tab.~\ref{tab:simpointpoly}. The average value of the
Polyakov loop in the adjoint representation can be computed by using the
relation 
\begin{equation}
\tr P^{\mathrm{adj}}=|\tr P|^2-1\ ,
\end{equation}
which is an easy consequence of Eq.~\eqref{eq:fundadj}, and from the discussion in
Sec.~\ref{sec:ren} if follows that $\langle \tr P^{\mathrm{adj}}\rangle$ scales to
the continuum as $\exp(c_a N_t)$.

Numerical results obtained for $\langle \tr P^{\mathrm{adj}}\rangle$ are shown
in Fig.~\ref{fig:Z}, in which some deviations from the asymptotic $\exp(c_aN_t)$
behaviour are also visible. To extract the value of $c_a$ fits of the form
\begin{equation}
\ln\langle P^{\mathrm{adj}}\rangle = k_0 + c_a N_t +\frac{k_1}{N_t}+\frac{k_2}{N_t^2}
\end{equation} 
have been performed, and the stability of the fit under changes of the fit
range and of the functional form adopted (i.e. by setting $k_2=0$) has been
investigated. As our final estimate we report the value
\begin{equation}
c_a=-0.45(2)\ . 
\end{equation}

\begin{figure}[b] 
\centering 
\includegraphics[width=0.9\columnwidth, clip]{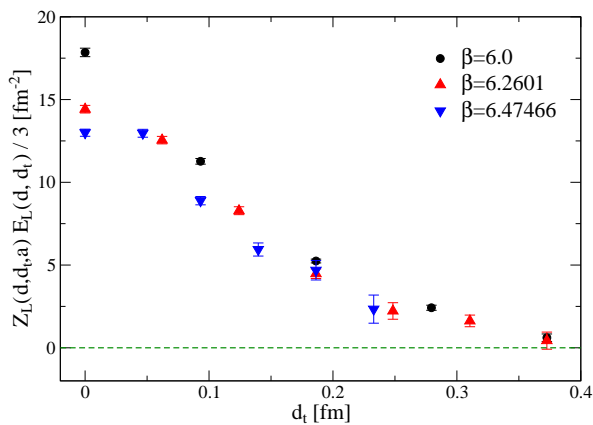}
\caption{Continuum scaling of $Z_L(d, d_t, a) E_L(d, d_t)$ for
$d_t\simeq 0.37\,\mathrm{fm}$. }
\label{fig:el_withz}
\end{figure}

Using this value for $c_a$ we can remove linear divergences from $E_L(d, d_t)$,
and in Fig.~\ref{fig:el_withz} the values of $Z_L(d,d_t,a)E_L(d, d_t)$ are
shown as a function of $d_t$ (in physical units). The lattice spacing
dependence of data in Fig.~\ref{fig:el_withz} is much milder than that observed
in Fig.~\ref{fig:el_noz}, however we have to remember that the renormalization
factor $Z_1(a,d_t)$, needed to take care of logarithmic divergences, is still
missing.  A consequence of this fact is that the scaling to the continuum of
points associated to distances $d_t=0$ and $d_t>0$ is different, as can be seen
in Fig.~\ref{fig:el_withz} and will be most clearly evident from the
considerations of the next section.

\subsection{Dual superconductivity parameters}\label{sec:dual}

According to the dual superconductor model of color confinement
\cite{Mandelstam:1974pi, Parisi:1974yh, tHooft:1975yol} the vacuum of
nonabelian gauge theories behaves as a ``dual'' superconductor, in which
condensation of chromomagnetic degrees of freedom produces a ``dual'' Meissner
effect, that squeezes the chromoelectric field lines into flux tubes producing
confinement. The characteristic feature of this model is to provide a
conceptually simple and physically appealing framework to interpret some
nonperturbative aspects of gauge theories. 

A generic feature of any superconductor (dual or not) is the presence of two
typical lengths in the infrared effective theory: the coherence length $\xi$
and the penetration length $\lambda$. The values of these lengths characterize
the functional form of the flux tube profile, with the penetration length being
associated to the exponential decrease of the field far from the center of the
flux tube. In order to determine both $\xi$ and $\lambda$ starting from the
data presented in the previous section, we will follow the approach first
adopted in \cite{Cea:2012qw} (and then used in \cite{Cea:2014uja, Cea:2015wjd,
Cea:2017ocq, Bonati:2018uwh}).

\begin{figure}[t] 
\centering 
\includegraphics[width=0.9\columnwidth, clip]{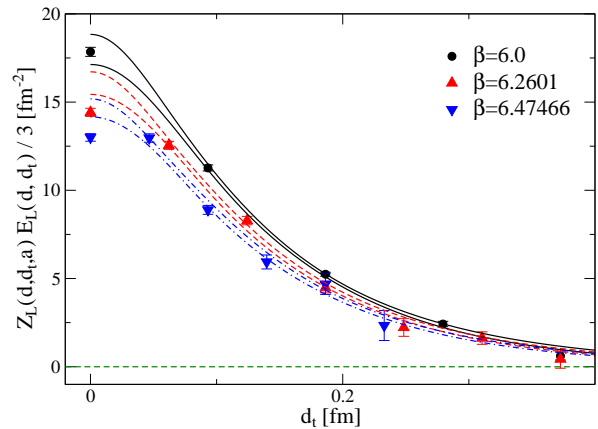}
\caption{Same data as in Fig.~\ref{fig:el_withz}, together with $1\sigma$
bands obtained from the combined fit described in the text. }
\label{fig:el_withz_globfit}
\end{figure}

In this approach the following parametrization of the longitudinal component of
chromoelectric field inside the flux tube is used:
\begin{equation}\label{eq:clem_fit}
E_L^{(R)}(d_t)=\frac{\phi}{2\pi}\frac{\mu^2}{\alpha}\frac{K_0\big(\sqrt{\mu^2 d_t^2+\alpha^2}~\big)}{K_1(\alpha)}\ ,
\end{equation} 
where $K_0$, $K_1$ are modified Bessel functions of the second kind, and $\phi,
\alpha$ and $\mu$ are fit parameters. This is the ``dual'' version of the
parametrization introduced in \cite{Clem} for the longitudinal magnetic field
inside a vortex line in type II superconductors. The parameter $\mu$ is just
the inverse of the penetration length, $\mu=1/\lambda$, while the relation
between the fit parameters in Eq.~\eqref{eq:clem_fit} and the coherence length
is less direct: it can be shown (see \cite{Clem}) that the
Ginzburg-Landau parameter $\kappa=\lambda/\xi$ is related to $\alpha$ by the
relation
\begin{equation}\label{eqkappa}
\kappa=\frac{\sqrt{2}}{\alpha}\sqrt{1-\frac{K_0^2(\alpha)}{K_1^2(\alpha)}}\ .
\end{equation}
Values of $\kappa$ smaller than $1/\sqrt{2}$ correspond to superconductors of type I, 
while $\kappa>1/\sqrt{2}$ for type II superconductors (see e.g. \cite{Tinkham}).

If we try to fit each of the fixed $\beta$ data sets shown in
Fig.~\ref{fig:el_withz} by using the parametrization in
Eq.~\eqref{eq:clem_fit}, we immediately realize that the quality of the fits
degrades as the coupling is increased. This is a consequence of the
previously noted fact that data points at $d_t=0$ and at $d_t>0$ scale to the
continuum in different ways, due to the different logarithmic divergences in
the two cases. If on the other hand we simply discard the point at $d_t=0$ from
each of our data sets, the precision of our data is not enough for the fit to
provide significant information on $\alpha$, and consequently on the
Ginzburg-Landau parameter $\kappa$. 

\begin{figure}[t] 
\centering 
\includegraphics[width=0.9\columnwidth, clip]{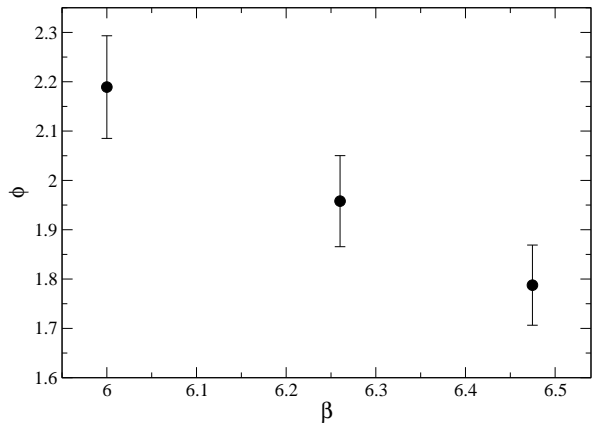}
\caption{Continuum scaling of the parameters $\phi$
obtained by using the combined fit described in the text.}
\label{fig:phis}
\end{figure}

We thus decided to perform a combined fit of all our data at $d_t>0$ keeping
three different $\phi$ parameters, corresponding to the three lattice spacings
used, since $\phi$ is sensitive to the multiplicative renormalization
$Z_1(a,d)$. Using $c_a=-0.45$ we obtain the best fit shown in
Fig.~\ref{fig:el_withz_globfit} and the final results for the superconductivity
parameters are
\begin{equation}\label{eq:sup_param}
\frac{1}{\lambda}=\mu=7.4(6)\mathrm{fm}^{-1},\quad \alpha=0.7(2), \quad \kappa=1.8(6)\ ,
\end{equation}
while the values obtained for the $\phi$s parameters shown in
Fig.~\ref{fig:phis}. The $\chi^2$ test for this fit gives 
$\chi^2/\mathrm{dof}=16/10$ that is somehow large but still acceptable, 
and the final results are almost unchanged also for $c_a=-0.47$
and $c_a=-0.43$, which means that the uncertainty in $c_a$ is not the
main source of error in our final results.

\section{Conclusions}\label{sec:concl}

In this paper we presented the results of our study of the longitudinal
chromoelectric component of the color flux tube, performed by using the
connected correlator. Measures were carried out by means of stochastically
exact techniques, without any smoothing, in order to investigate the
possibility of using the connected correlator $\rho_{conn}$ in a coherent
field-theoretical setup. 

We first investigated the renormalization properties of $\rho_{conn}$, showing
that it is multiplicatively renormalizable, and reducing the problem of its
renormalization to the determination of three renormalization constants. One of
these constants (denoted by $c_a$) is related to linear divergences, while the
other two take care of the logarithmic divergences in the cases $d_t=0$ and
$d_t\neq 0$ respectively. 

We then fixed the value of $c_a$, by studying the $\beta$ dependence (at fixed
temperature) of the Polyakov loop in the adjoint representation. Using the
value of $c_a$ obtained in this way ($c_a=-0.45(2)$), we removed linear
divergences from $\rho_{conn}$, obtaining the results shown in
Fig.~\ref{fig:el_withz}. While it is important to stress that these data are
still not renormalized (since logarithmic divergences have not been removed),
it is possible to extract from them quantities of direct
physical interest. 

In particular, starting from the functional form of the longitudinal
chromoelectric field, we evaluated the coherence length $\xi$ and the
penetration length $\lambda$ of the dual superconductor model. The numerical
values of these quantities, extracted using a flux tube of length $d\simeq
0.37\,\mathrm{fm}$, are reported in Eq.~\eqref{eq:sup_param}.  The
values of $\xi$, $\lambda$ and $\kappa$ reported in the literature have been
obtained using quite larger $d$ values, so that a direct comparison can not be
performed.  It is nevertheless interesting to note that our estimate of the
penetration length $\lambda$ is in good agreement with previous determinations.
Our Ginzburg-Landau parameter (and consequently our coherence length) is
instead quite different from the one obtained in similar studies carried out by
using smoothing \cite{Cea:2012qw, Cea:2014uja} (where $\kappa\approx 0.2$ was
found), being closer to older results suggesting the $SU(2)$ vacuum to be at
the boundary between type I and type II superconductivity \cite{Bali:1997cp,
Gubarev:1999yp, Koma:2003hv, Haymaker:2005py, Chernodub:2005gz,
DAlessandro:2006hfn}. 

This result suggests that smoothing could introduce some systematics in the
determination of the flux tube profile, and we get the following intuitive
picture: $\xi$ is the typical scale of the bulk of the flux tube, which
broadens under smoothing, while $\lambda$ is related to the large distance
behaviour of the tails, which is almost unaffected by smoothing. As a consequence
we expect smoothing to leave almost unaltered $\lambda$ and to decrease
$\kappa=\lambda/\xi$.

While this picture seems appealing we also have to keep in mind the limitations
of our computation: first of all our determination of $\kappa$ has a $30\%$
relative error, so this effect could just be a statistical fluctuation.
Moreover the distance between the Polyakov loops used to extract $\xi$ and $\lambda$
was only about $0.37\,\mathrm{fm}$, which is surely not asymptotically large;
as a consequence a contamination from the Coulomb component of the flux tube is
possible (see \cite{Baker:2018mhw} for a discussion on this point). 

As noted before, most of the results reported in the literature adopt quite
larger values of $d$ to extract $\lambda$ and $\xi$, so that a fair comparison
with our results Eq.~\eqref{eq:sup_param} is not possible. However a direct
comparison can be made between Fig.~\ref{fig:phis} above and Fig.~2 of
\cite{Cea:2014uja}, where the flux tube profile is reported for the case
$\beta=6.0$ and $d=4a$ (computed by using a $20^4$ lattice): the half-width at
half-maximum of the flux tube in Fig.~\ref{fig:phis} (for $\beta=6.0$) is about
$0.12\,\mathrm{fm}$, while the corresponding value extracted from Fig.~2 of
\cite{Cea:2014uja} is about $0.23\,\mathrm{fm}$. This is consistent with the
possibility that smoothing increases the thickness of the flux tube.

A more complete investigation of the long distance structure of the flux
tube, performed by using higher statistics and larger values of the distance
between the Polyakov loops, is surely matter for further studies, just as the
determination of the renormalization constants associated to the logarithmic
diverges of $\rho_{conn}$.

\acknowledgments

It is a pleasure to thank Massimo D'Elia and Alessandro Papa for useful comments.
Numerical simulations have been performed on the CSN4 cluster of the Scientific
Computing Center at INFN-PISA, and on the MARCONI machine at CINECA, based on
the agreement between INFN and CINECA (under project INF19\_npqcd).

\end{document}